\documentclass[12pt]{iopart}

\usepackage{iopams} 
\usepackage{setstack}
\usepackage{mathrsfs}
\usepackage{epsfig}
\usepackage{graphicx}
\usepackage{bm}

\begin{document}

\newcommand{\ped}[1]{_{\mathrm{#1}}}
\newcommand{\api}[1]{^{\mathrm{#1}}}
\newcommand{\diff}[1]{\mathrm{d}#1}

\title{All-optical trapping and acceleration of heavy particles}

\author{F Peano$^1$, J Vieira$^1$, L O Silva$^1$, R Mulas$^2$ and G Coppa$^2$}
\address{$^1$ GoLP/Institudo de Plasmas e Fus\~ao Nuclear, Instituto Superior T\'ecnico, 1049-001 Lisboa, Portugal}
\address{$^2$ Dipartimento di Energetica, Politecnico di Torino, 10129 Torino, Italy}
\eads{\mailto{fabio.peano@ist.utl.pt}, \mailto{luis.silva@ist.utl.pt}}

\begin{abstract}
A scheme for fast, compact, and controllable acceleration of heavy particles in vacuum is proposed, in which two counterpropagating lasers with variable frequencies drive a beat-wave structure with variable phase velocity, thus allowing for trapping and acceleration of heavy particles, such as ions or muons. Fine control over the energy distribution and the total charge of the beam is obtained via tuning of the frequency variation. The acceleration scheme is described with a one-dimensional theory, providing the general conditions for trapping and scaling laws for the relevant features of the particle beam. Two-dimensional, electromagnetic particle-in-cell simulations confirm the validity and the robustness of the physical mechanism.
\end{abstract}


\submitto{\NJP}
\maketitle

\section{Introduction}
\label{sec:intro}
Compact acceleration of charged particles has become a central research topic due to the recent developments in the technology of ultraintense lasers in the infrared (IR) optical regime. So far, the main research directions in the field have been electron acceleration in laser-driven plasma waves \cite{wakefield_1}--\cite{wakefield_5}, and proton or light-ion acceleration in laser-solid interactions (cf. \cite{ion_accel_0}--\cite{Esirkepov} and, for a review, \cite{Mendonca,Bulanov}), which is of particular interest for possible applications to the treatment of tumors with hadron beams \cite{Kraft}-\cite{ELI}.
These acceleration schemes are indirect, in the sense that the acceleration is provided by laser-driven field structures in plasmas. In the case of electrons, the availability of relativistic radiation intensities ($I\gtrsim10^{18}$ W/cm$^2$ with IR lasers) also motivated a number of proposals for direct, all-optical acceleration in vacuum by the electromagnetic (EM) field of one or more laser pulses  \cite{Esarey_1}--\cite{vacuum_8}; in the case of ions, for which relativistic radiation intensities ($I\gtrsim10^{24}$ W/cm$^2$ with IR lasers) are still beyond the limits of current technology, direct acceleration has been mostly unexplored (only very recently, the possibility of direct ion acceleration by tightly focused, radially polarized lasers has been investigated \cite{Salamin_OL}).

In this paper, a new concept for direct, all-optical acceleration in vacuum is presented, which allows for fast, compact acceleration of heavy particles, including protons, heavy ions, and muons. The technique exploits two counterpropagating EM waves with variable frequency in order to drive a beat-wave structure with controllable phase velocity. If the frequency variation is chosen appropriately, the ponderomotive potential associated with the beat wave provides longitudinal trapping and long-term particle acceleration.
The main characteristics of the method are the following: (i) it can work with ``modest'', nonrelativistic radiation intensities (e.g., $I \ll 10^{24}$ W/cm$^2$ with IR lasers), provided that the available electromagnetic energy is sufficiently high (typically, $\gtrsim 100$ J); (ii) it does not involve any complex laser-matter coupling; (iii) it exploits an extremely simple and robust physical mechanism, operating in a test-particle regime, as a conventional accelerator, offering the possibility of direct, efficient control over the relevant beam features (mean energy, energy spread, and total accelerated charge) via regulation of the fundamental laser parameters (intensity, duration, and frequency variation); (iv) it can be adapted to any source of charged particles (including beams, and tenuous gases and plasmas), serving either as a pre-accelerator, creating an energetic beam from heavy particles with low energy, or as an energy booster, accelerating particles from a beam with a given energy.

The main advantages of the acceleration technique, controllability, robustness, and simplicity of the physical scheme, are obtained through two technological requirements \cite{Peano_IEEE}: (i) the method typically requires long, energetic laser pulses to maintain the accelerating structure for sufficiently long times; (ii) the method typically requires significant frequency excursions to achieve high energy gains. If these technological requirements can be met, the technique can represent an alternative option whenever beam quality (e.g., a low energy spread) and a high degree of control on the particle beam are critical (the pertinence of the present technique with respect to other laser-based techniques depends on the specific application). Furthermore, the present acceleration scheme could also be an option in applications for which no other laser-based acceleration techniques are currently available \cite{Bingham} (e.g., ultrafast acceleration of muons during, or immediately after, the muon cooling process in a muon collider or a neutrino factory machine \cite{NFMC1}--\cite{NFMC8}).   

The Paper is organized as follows: in Section \ref{sec:theory}, the basic physical mechanism of the acceleration scheme is described with a one-dimensional theory; in Section \ref{sec:trapping}, the resonant solutions of the system and the conditions for particle trapping are analyzed; in Section \ref{sec:scalings}, basic scaling laws for the properties of the particle beam are provided; in Section \ref{sec:simulations}, numerical results from self-consistent simulations in two dimensions are presented. Finally, in Section \ref{sec:conclusions}, the conclusions are stated.

\section{Physical mechanism}
\label{sec:theory}
The essential aspects of the method are illustrated with a one-dimensional, single-particle theory, in which the amplitude of both lasers is constant, and only one of the beams has variable frequency. For the sake of simplicity, the Lorentz reference frame where the initial velocity of the particle is zero, and the laser frequencies at $x=0$ are initially the same, is adopted. The first laser beam, propagating from left to right in the $x$ direction, is described by the vector potential ${\bf A}_1 = A_1\{\cos(\theta)\sin[k_0\xi+\phi(\xi)]\hat{{\bf e}}_y + \sin(\theta)\cos[k_0\xi+\phi(\xi)]\hat{{\bf e}}_z\}$, where $\theta$ determines the polarization type (e.g., $\theta = \pi/2$ for linear polarization along $z$, $\theta = \pi/4$ for circular polarization), $\xi = x - ct$, $k_0=2\pi/\lambda_0$ is the wavenumber at $\xi = 0$, and $\phi$ is an arbitrary function, such that its derivative $\phi^\prime$ vanishes for $\xi=0$.
The second beam propagates in the opposite direction, with vector potential ${\bf A}_2 = A_2\{\cos(\theta)\sin[k_0(-x-ct)]\hat{{\bf e}}_y + \sin(\theta)\cos[k_0(-x-ct)]\hat{{\bf e}}_z\}$. 
By resorting to the conservation of the transverse canonical momentum, the $x$ component of the equation of motion for the particle is written as
\begin{equation}
\frac{\diff{p_x}}{\diff{t}} = -\frac{q^2}{2\gamma Mc^2}\frac{\partial}{\partial x}({\bf A}_1+{\bf A}_2)^2 \mathrm{,}
\label{eq:x_motion}
\end{equation}
where $q$ and $M$ are the particle charge and mass, respectively, and $\gamma$ is the Lorentz factor. If $\hat{A}_j = \frac{qA_j}{Mc^2}\ll1$, for $j=1,2$, the longitudinal displacement due to the quiver motion can be neglected (here, $\hat{A}_j=\frac{qm}{eM}a_j$, where $m$ and $e$ are the electron mass and the elementary charge, and $a_j=\frac{eA_j}{mc^2}$ is the usual normalized vector potential, corresponding to the peak transverse momentum of an electron in the laser field). 
Hence, as long as the ratio between the low and high beat-wave frequencies is negligible along the particle trajectory $x\ped{p}(t)$ -- i.e., as long as $\left|\omega\ped{p}/\Omega\ped{p} \right|\ll 1$, where $\omega\ped{p} = (c-v\ped{p})\frac{\phi^{\prime}}{2}-k_0v\ped{p}$ and $\Omega\ped{p} = (c-v\ped{p})\frac{\phi^{\prime}}{2}+k_0c$, with $v\ped{p}=\frac{\diff{x\ped{p}}}{\diff{t}}$ -- the equation of motion can be averaged over the fast time scale to yield, in dimensionless form ($\hat{t} = k_0ct$, $\hat{x} = k_0x$, and $\hat{p}_x = \frac{p_x}{Mc}$),
\begin{equation}
\frac{\diff{\hat{p}_x}}{\diff{\hat{t}}} = -\frac{\hat{A}_1\hat{A}_2}{2\gamma}\frac{\partial}{\partial \hat{x}}\cos\left[2\hat{x}+\phi(\hat{x}-\hat{t})\right] \mathrm{,}
\label{eq:x_motion_av}
\end{equation}
where $\gamma = \{1+\hat{p}_x^2+\hat{A}_1^2 /2+\hat{A}_2^2 /2+\hat{A}_1\hat{A}_2\cos[2\hat{x}+\phi(\hat{x}-\hat{t})]\}^{1/2}$ now denotes the average Lorentz factor \cite{Mora_Antonsen}. 
Equation \eref{eq:x_motion_av} can be obtained from the Hamiltonian $\mathscr{H}\left(\hat{x},\hat{p}_x,\hat{t}\right)=\{1+\hat{p}_x^2+\hat{A}_1^2 /2+\hat{A}_2^2 /2+\hat{A}_1\hat{A}_2\cos[2\hat{x}+\phi(\hat{x}-\hat{t})]\}^{1/2}$. Hence, the variation of the particle energy, $\frac{\diff{\gamma}}{\diff{\hat{t}}}=\frac{\partial\mathscr{H}}{\partial \hat{t}}$, is given by
\begin{equation}
\frac{\diff{\gamma}}{\diff{\hat{t}}} = \frac{\hat{A}_1\hat{A}_2}{2\gamma} \frac{\partial}{\partial \hat{t}}\cos[2\hat{x}+\phi(\hat{x}-\hat{t})] \mathrm{.}
\label{eq:energy}
\end{equation}
If there is no frequency variation ($\phi^{\prime} = 0$), $\mathscr{H}$ is time independent and $\gamma$ remain constant (in agreement with the Lawson-Woodward theorem \cite{Esarey_1,LW_1,LW_2}). In the presence of frequency variations, the phase velocity of the ponderomotive beat wave varies in time, allowing for wave-particle energy transfer. If $\phi$ is chosen appropriately, the particle is trapped by the accelerating beat wave: since, for a trapped particle, the condition $\left|\omega\ped{p}/\Omega\ped{p} \right|\ll 1$ is always satisfied and Eq. \eref{eq:x_motion_av} is valid for every $\hat{t}$, the particle can be accelerated continuously and over long times.

\section{Phase-locking and particle trapping}
\label{sec:trapping}
For a particular choice of the function $\phi$, equation \eref{eq:x_motion_av} admits a resonant solution, $\hat{X}(\hat{t})$, such that $2\hat{X}(\hat{t})+\phi[\hat{X}(\hat{t})-\hat{t}]=\phi_0$. In this solution, the particle is exactly phase-locked to the beat wave, i.e., its trajectory is identical to the trajectory of the point of the beat wave having constant phase $\phi_0$, here denoted with $\hat{x}_{\phi_0}(\hat{t})$.
The resonance occurs when $\phi$ depends on $\hat{\xi}=\hat{x}-\hat{t}$ as
\begin{equation}
\phi(\hat{\xi}) = \phi_0 -\hat{\xi} +\frac{1}{2\mu}\log(1+2\mu\hat{\xi})\mathrm{,}
\label{eq:phi_exact}
\end{equation}
where $\mu = \hat{A}_1\hat{A}_2\sin(\phi_0)/\gamma_{0_\perp}^2$, with $\gamma_{0_\perp}^2=1+\hat{A}_1^2/2+\hat{A}_2^2/2+\hat{A}_1\hat{A}_2\cos(\phi_0)$. The resonant trajectory can be expressed in parametric form, as a function of the proper time $\tau$ (defined as $\diff{\tau}/\diff{\hat{t}}=\gamma^{-1}$), as
\numparts
\begin{eqnarray}
\hat{X}(\tau) &=&\frac{1}{2\mu}[\mu_0\tau(\mu_0\tau/2-1)-\log(1-\mu_0\tau)] \mathrm{,}
\label{eq:res_a}\\
\hat{t}(\tau) &=& \frac{1}{2\mu}[\mu_0\tau(1-\mu_0\tau/2)-\log(1-\mu_0\tau)] \mathrm{,}
\label{eq:res_b}
\end{eqnarray}
\endnumparts
with $\mu_0 = \gamma_{0_\perp}\mu$. Accordingly, the particle energy increases as $\gamma(\tau)=\gamma_{0_\perp}[1+(1-\mu_0\tau)^2]/[2(1-\mu_0\tau)]$. 
For $\hat{t}\ll\mu^{-1}$, the energy grows quadratically in time as $\gamma_{0_\perp}(1+\mu^2\hat{t}^2/2)$. For $\hat{t}\gg\mu^{-1}$, the energy scaling depends on the sign of $\mu$:
if $\mu>0$ (the particle and the variable-frequency laser are copropagating), the frequency must be increased to maintain phase-locking and the energy grows exponentially as $\gamma_{0_\perp}/2 \ \exp(2\mu\hat{t}-1/2)$; if $\mu<0$, the laser frequency must be decreased and the energy scales with the power law $\gamma_{0_\perp}(|\mu|\hat{t})^{1/2}$.

\begin{figure}[!htbp]
\centering \epsfig{file=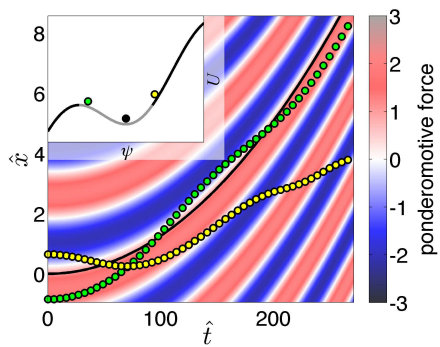, width=2.5in}
\caption{Space-time evolution of the dimensionless ponderomotive force $-\frac{\partial}{\partial{\hat{x}}}\cos[2\hat{x}+\phi(\hat{x}-\hat{t})]$ (color scale, red and blue denoting accelerating and decelerating regions, respectively), and trajectories of a resonant particle (solid line), a non-resonant, trapped particle (green markers), and a non-resonant, untrapped particle (yellow markers), for a case where $\phi$ is given by Equation \eref{eq:phi_exact}, $\hat{A}_1\hat{A}_2=5\times10^{-4}$, and $\phi_0=5\pi/6$. The inset represents the effective potential $U$ at $\hat{t}=0$, gray marking the region where trapping is possible, and the corresponding initial positions of the three particles.}
\label{fig:scheme}
\end{figure}

In order to determine the condition under which the acceleration mechanism is effective, the stability of the particle motion has been investigated. The analysis reveals that stable solutions, and, consequently, the occurrence of particle trapping, exist whenever the frequency variation is sufficiently slow, independently of the particular shape of $\phi$.
Expressing Eq. \eref{eq:x_motion_av} in terms of the phase difference between the particle and the wave, $\psi=2[\hat{x}-\hat{x}_{\phi_0}(\hat{t})]$, one obtains
\begin{equation}
\frac{\diff^2{\psi}}{\diff{\tau^2}} = -\frac{\partial}{\partial\psi}U(\psi,\tau) \mathrm{,}
\label{eq:dpsi_dtau}
\end{equation}
where
\begin{equation} 
U(\psi,\tau) = 2\hat{A}_1\hat{A}_2 \cos\Big\{ \psi+2\hat{x}_{\phi_0}\big[\hat{t}(\tau)\big] + \phi\big[\hat{\xi}_{\phi_0}(\tau)+\psi/2\big]\Big\} + 2\alpha_{\phi_0}(\tau)\psi \mathrm{,}
\label{eq:U}
\end{equation} 
with $\hat{\xi}_{\phi_0}(\tau)=\hat{x}_{\phi_0}[\hat{t}(\tau)]-\hat{t}(\tau)$ and $\alpha_{\phi_0}(\tau)=\frac{\diff^2}{\diff{\tau^2}}\hat{x}_{\phi_0}[\hat{t}(\tau)]$. The quantity $U$ plays the role of an effective potential and its shape determines the possible occurrence of particle trapping, which is allowed only if the frequency variation is smooth enough to guarantee that $U$ presents local minima. For trapped particles, $|\psi|\ll2\hat{\xi}_{\phi_0}$ and $U$ can be approximated as $2\hat{A}_1\hat{A}_2 \cos\{[1+\phi^{\prime}(\hat{\xi}_{\phi_0})/2]\psi +\phi_0\} + 2\alpha_{\phi_0}\psi$, leading to the necessary condition for particle trapping
\begin{equation}
\left|\alpha_{\phi_0}\right|<\alpha_{\scriptscriptstyle\mathrm{M}}=\hat{A}_1\hat{A}_2\left[1+\phi^{\prime}\left(\hat{\xi}_{\phi_0}\right)/2\right] \mathrm{,}
\label{eq:cond}
\end{equation}
where $\alpha_{\scriptscriptstyle\mathrm{M}}$ is the maximum value of the ponderomotive force. The physical interpretation of Equation \eref{eq:cond} is that trapping is allowed only if the maximum ponderomotive force of the beat wave is greater than the inertial force associated with the beat-wave acceleration.
The limit in which $|\alpha_{\phi_0}|=\alpha_{\scriptscriptstyle\mathrm{M}}$ corresponds to the resonant solution of Equations \eref{eq:res_a} and \eref{eq:res_b}, in the particular case $\phi_0 = \pm \pi/2$. When $\phi$ is given by \eref{eq:phi_exact} and $|\phi_0| \neq \pi/2$, regions where trapping is possible always exist, although the solution is stable for $\cos(\phi_0)<0$ and unstable for $\cos(\phi_0)>0$, with $\psi=0$ corresponding to the bottom and the top of the potential well, respectively. 
The initial relative width of the potential well, $\frac{\Delta\psi}{2\pi}\simeq 1-\frac{2}{\pi}\arcsin[\alpha_{\phi_0}(0)/\alpha_{\scriptscriptstyle\mathrm{M}}]$, provides an estimate for the trapping efficiency.
As an example, the typical trajectories of three test particles are shown in Fig. \ref{fig:scheme}, along with the evolution of the beat-wave ponderomotive force. 

The generality of the trapping condition guarantees that the resonance criterion \eref{eq:phi_exact} can be relaxed, provided that $\phi$ is sufficiently smooth, opening the possibility of controlling the final particle energy and the total accelerated charge by tuning the frequency variation.
An important choice for $\phi$, particularly relevant in experiments, is $\phi(\hat{\xi}) = \phi_0 + \sigma\hat{\xi}^2$, corresponding to a linear chirp. In this case, the beat-wave trajectory, $\hat{x}_{\phi_0}(\hat{t}) = \hat{t} - [1 - (1-2\sigma\hat{t})^{1/2}]/\sigma$, is such that $\alpha_{\phi_0}(\hat{t}) = -\sigma(1-2\sigma\hat{t})^{1/2}/[1-2(1-2\sigma\hat{t})^{1/2}]^2$. For $\hat{t}=0$, trapping regions exist if $|\sigma| < \hat{A}_1\hat{A}_2$.
The energy of the trapped particles grows approximately as $[2(1-2\sigma\hat{t})^{-1/2}-(1-2\sigma\hat{t})^{-1}]^{-1/2}$. Depending on the sign of $\sigma$, the trapping region evolves in time differently: if $\sigma<0$ (trapped particles move from left to right), it becomes wider (cf., movie mov1 in \cite{movies}); if $\sigma>0$ (trapped particles move from right to left), it becomes narrower, until disappearing. Thus, long-term stable acceleration can be achieved only when the accelerating particles and the chirped laser are copropagating.

\section{Estimates for the basic beam properties}
\label{sec:scalings}
In the nonrelativistic regime, the maximum energy gain, $\Delta \epsilon_{\scriptscriptstyle\mathrm{M}}$, can be estimated as $\Delta \epsilon_{\scriptscriptstyle\mathrm{M}}[\mathrm{MeV}]\approx 0.8$ $q^4[e]$ $M^{-3}$[amu] $ I_1[10^{20}\mathrm{W/cm^2}]$ $I_2[10^{20}\mathrm{W/cm^2}]$ $\lambda_0^2[\mu\mathrm{m}]$ $\Delta T^2$[ps]; the corresponding acceleration distance scales as $\Delta x[\mu\mathrm{m}] \approx 6$ $q^2[e]$ $M^{-2}$[amu] $I_1^{1/2}[10^{20}\mathrm{W/cm^2}]$ $I_2^{1/2}[10^{20}\mathrm{W/cm^2}]$ $\lambda_0[\mu\mathrm{m}]$ $\Delta T^2$[ps].
By varying the chirp rate $\sigma$, the energy gain $\Delta \epsilon$ can be controlled, according to $\Delta \epsilon = (\sigma/\sigma_{\scriptscriptstyle\mathrm{M}})^2\Delta \epsilon_{\scriptscriptstyle\mathrm{M}}$. Since the trapping efficiency decreases when increasing $|\sigma|$, the chirp rate must be chosen in order to obtain the desired trade-off between energy gain and total accelerated charge.

By analyzing the range of validity of the model, conclusions can be drawn for more general situations, in which spatial distributions of charged particles are considered and the effects due to the finite-size of the laser pulses are taken into account.
In the presence of a spatial charge distribution, the scheme remains effective if the particle density $n$ is low enough that the space-charge field does not perturb the EM beat-wave structure. Assuming that a single particle species is present, this establishes the criterion $n[10^{19}\mathrm{cm}^{-3}] \ll q^{-1}[e]I_1^{1/2}[10^{20}\mathrm{W/cm^2}]I_2^{1/2}[10^{20}\mathrm{W/cm^2}]$. When taking into account the finite size of the laser beams,
both the longitudinal and the transverse dependence of the laser envelope must be considered.
Since long pulses (e.g., with duration in the range 1 ps - 1 ns) are required to obtain significant energy gain, $\hat{A}_1$ and $\hat{A}_2$ can be considered as slowly varying functions of $\hat{t}$. 
Hence, intensities high enough for particle trapping exist only for a fraction of the overlap between the laser pulses, and the duration of the acceleration process is limited to a fraction of the pulse duration, determined by the specific chirp law.
For similar reasons, the transverse cross section of the accelerating region is on the order of $\pi W_0^2$, where $W_0$ is the overlapped spot size of the lasers, and the length of the acceleration process is always limited to distances on the order of the Rayleigh length, $Z\ped{R} = \pi W_0^2/\lambda_0$.
Therefore, the total accelerated charge $Q$ scales as $Q\mathrm{[pC]} \approx 16 \eta\ped{tr}q[e]$ $n[10^{19}\mathrm{cm}^{-3}]$ $W_0^4[\mu\mathrm{m}]/\lambda_0[\mu\mathrm{m}]$, where $\eta\ped{tr}$ is the fraction of particles trapped by the beat wave.
The finite transverse size of the lasers also causes a slow outward drift due to the transverse ponderomotive force of the lasers. If the intensities of the two lasers are comparable, the ratio between transverse and longitudinal momenta scales as $\frac{\lambda_0}{2\pi W_0}\ll1$. Transverse focusing and trapping can be obtained using laser pulses with appropriate shape \cite{Kawata}.

The above scalings indicate that the acceleration scheme described here does allow for compact, fast acceleration of heavy particles.
In the case of protons, proof-of-principle experiments producing monoenergetic beams in the 100 keV - 1 MeV energy range could already be performed using kJ-level, suitably chirped pulses, with frequency excursions on the order of 5-10\% of the central frequency (cf. \cite{Peano_IEEE}), which are potentially available with current laser technology. 
Furthermore, the technique could open the way to ultrafast acceleration of low-energy muons \cite{Bingham}, for which multi-MeV energy gains, on a ps time scale, could be achieved using lasers with moderate energy (on the order of 1 J) \cite{Peano_talk}; such a possibility could be critical to minimize muon-decay losses.

\section{Numerical simulations}
\label{sec:simulations}
\begin{figure}[!htb]
\centering \epsfig{file=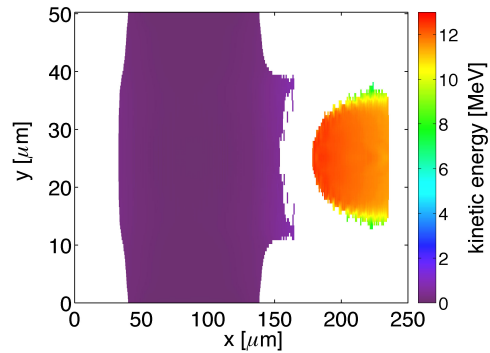, width=2.5in}
\caption{(Color online) Mean kinetic energy of the protons as a function of position, after $6$ ps.}
\label{fig:den}
\end{figure}

The effectiveness of the method has been confirmed with EM particle-in-cell simulations, using the OSIRIS 2.0 framework \cite{osiris}, accounting self-consistently for space-charge and propagation effects.
Here, a selection of results is presented from a two-dimensional (2D) simulation, in which the source of particles to be accelerated is provided simply by a slab of hydrogen, with thickness 75 $\mu$m and proton density $n = 5\times10^{16}$ cm$^{-3}$. The computational domain is rectangular [$L_x=300$ $\mu$m ($5\times10^4$ cells) and $L_y=50$ $\mu$m ($100$ cells)], with open-space boundary conditions and periodic boundary conditions in the $x$ and $y$ directions, respectively; the left face of the hydrogen slab is located at $x=40$ $\mu$m. Two approximately gaussian laser pulses, with spot size $W_0=10$ $\mu$m and duration $4.2$ ps, are focused on the mid plane of the slab. The pulse propagating from right to left has peak intensity $I_2 = 8.5\times10^{20}$ W/cm$^2$, fixed frequency, and central wavelength $\lambda_0=800$ nm. The pulse propagating from left to right has peak intensity $I_1 = 1.3\times10^{21}$ W/cm$^2$ and its phase varies quadratically as $k_0\xi + \sigma(\xi-\xi_0)^2$, where $\xi$ indicates the distance from the pulse center, $\sigma=-2\times10^{-5} \ k_0^2$ is the chirp coefficient, and $\xi_0=6.25\times10^{3} \ k_0^{-1}$ is the point of the pulse where $k=k_0$, which need to be specified when using finite-length pulses. Field ionization of the neutral atoms \cite{Samuel} is calculated by using the Ammosov-Delone-Krainov rates \cite{ADK}. The laser pulses are launched from the opposite $x$ boundaries of the computational domain and hit the left face of the hydrogen slab at the same time.
\begin{figure}[!htb]
\centering \epsfig{file=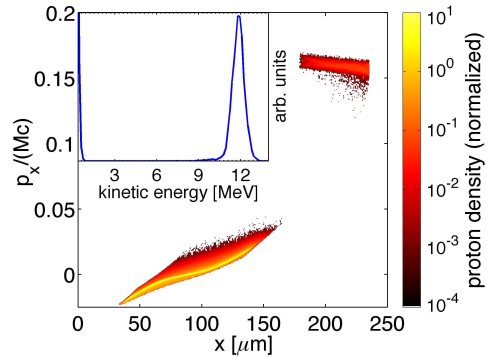, width=2.5in}
\caption{(Color online) Proton distribution in the $x$-$p_x$ phase space, after $6$ ps. The inset shows the corresponding kinetic-energy spectrum.}
\label{fig:phase}
\end{figure}
After $1.8$ ps, the trapping process begins and 4.2\% of the protons initially within the focal region of the lasers are extracted from the bulk distribution and accelerated (cf., movie mov2 in Ref. \cite{movies}). As the acceleration continues, these protons form a bunch that separates from the main distribution both in space and in momentum, as illustrated in Figs. \ref{fig:phase} and \ref{fig:den}. After $6$ ps, the proton beam has a mean energy of $12$ MeV with a $7.5$\% energy spread, approximately $60$ $\mu$m long and $20$ $\mu$m wide, with a transverse momentum spread on the order of $10^{-2}$ $Mc$.
Much lower energy spreads, down to 1\% and below, can be obtained by employing longer pulses with the same energy and lower intensities.
The expansion of the remaining plasma bulk, visible in Fig. \ref{fig:phase} and in mov2 in \cite{movies}, is driven by the space charge that forms because of the rapid expansion of the hot electrons produced by the lasers (for the radiation intensities considered  here, the electron dynamics in the beat wave is dominated by stochastic heating \cite{heating_1,heating_2}).
These results confirm the potential of the proposed technique for direct acceleration as a method for producing high-quality beams of heavy particles in a controlled way. 
Other simulation results, to be presented in future works, suggest that, for higher plasma densities, the space-charge field can affect the dynamics and, under appropriate conditions, it can even improve both trapping and acceleration.

\section{Conclusions}
\label{sec:conclusions}
Trapping and acceleration of heavy particles in a purely optical structure, generated by two counterpropagating laser beams with variable frequency, has been demonstrated. Relying on an extremely simple and robust physical mechanism, the method proposed here offers unique capabilities of detailed control over the main features of the particle beam. This makes the new acceleration technique a promising candidate for any application where the controllable production of high-quality beams of heavy particles in compact machines is critical, and for applications where ultrafast acceleration is crucial (e.g., the production of muon beams in a muon collider or a neutrino factory machine) \cite{Peano}.

\ack
Work partially supported by FCT (Portugal) through grants POCI/FIS/55095/2003 and SFRH/BD/22059/2005, and by the European Community - New and Emerging Science and Technology Activity under the FP6 "Structuring the European Research Area" programme (project EuroLEAP, contract number 028514).
The authors would like to acknowledge Prof. Tom Katsouleas for stimulating discussions, Prof. Robert Bingham for discussions and for suggesting the possibility to accelerate muons with the scheme described here, Prof. Ricardo Fonseca and Michael Marti for discussions and for help with the OSIRIS simulations, performed at the expp and IST clusters in Lisbon.

\end{document}